\documentstyle[11pt]{article}
\def\lsim{\mathrel{\rlap{\lower 4pt \hbox{\hskip 1pt $\sim$}}\raise 1pt \hbox
	{$<$}}}
\def\gsim{\mathrel{\rlap{\lower 4pt \hbox{\hskip 1pt $\sim$}}\raise 1pt \hbox
	{$>$}}}

\def\o{\over}
\begin{document}
\begin{titlepage}
\begin{flushright}
LPTHE/95-26,  SUSX-TH/95-31, December 1995 \\
%(\today)\\
\end{flushright}
\begin{center}
\LARGE
{\bf Quantum  bosonic string energy-momentum tensor 
  in Minkowski space-time.}\\
\vspace{.6cm}
\normalsize
\large{E.\@ J.\@ ~Copeland$^{\dag}$, H.\@ ~J.\@ ~de~Vega$^{\dag\dag}$\\ 
and\\ A.\@ V\'{a}zquez$^{\dag}$} \\
\normalsize   
\vspace{.4 cm} 
$^{\dag}${\em School of Mathematical and Physical Sciences, \\
University of Sussex, \\ Falmer, Brighton BN1 9QH,~U.~K. \\
\vspace{.3cm}
$^{\dag\dag}$LPTHE,\footnote{Laboratoire Associ\'{e} au CNRS UA 280.}
Universit\'{e} P. et M. Curie  (Paris VI) 
et  Universit\'e D. Diderot (Paris VII), Tour 16,  $1^{er}$
\'{e}tage, 4,
place Jussieu, 75252 Paris Cedex 05, France.}\\
\vspace{.3 cm}
\end{center}
\baselineskip=24pt
\begin{abstract}
\noindent \\
The quantum energy-momentum tensor  ${\hat T}^{\mu\nu}(x)$ 
is computed for strings in
Minkowski space-time. We compute its expectation value for different
physical string states both for open and closed bosonic strings. 
The states considered are described by normalizable wave-packets in
the center of mass coordinates. We find in particular that ${\hat 
T}^{\mu\nu}(x)$ is {\bf finite} which could imply that the classical
divergence  
that occurs in string theory as we approach the string position is removed at 
the quantum level as the string position is smeared out by quantum
fluctuations. 
 \noindent \\
For massive string states the expectation value of 
${\hat T}^{\mu\nu}(x)$ vanishes at leading order (genus zero).
\ For massless string states it has a non-vanishing value which we
explicitly compute and analyze for both spherically and cylindrically
symmetric  wave packets. The energy-momentum tensor components propagate
outwards as a massless lump peaked at $ r = t $.

\end{abstract}
\vspace{-.3cm}
\end{titlepage}
\section{Introduction.}
\hspace*{.9cm} String theory has emerged as the most promising
candidate to reconcile general relativity with quantum mechanics and
unify gravity with the other fundamental interactions.\ It makes sense
therefore to investigate the gravitational consequences of strings as we
approach the Planck scale.

When particles scatter at energies of the order of or larger than the
Planck mass, the interaction that dominates their collision is the
gravitational one, at these energies the picture of particle fields or
strings in flat space-time ceases to be valid, the curved space-time
geometry created by the particles has to be taken into account.\ This
has been our motivation to investigate the possible gravitational
effects arising from an isolated quantum bosonic string living in a flat
space-time background, so we may begin a study of the
scattering process of strings merely by the gravitational interaction
between them.\ 
The systematic study of quantum strings in physically relevant curved
space-times was started in \cite{dvs} and is reviewed {\it in extenso} in
ref.\cite{eri}.

In this paper, we calculate the energy-momentum tensor of both closed and open
quantum bosonic strings in $3+1$ dimensions. Our target space is the
direct product of four dimensional Minkowski space-time times a compact
manifold taking care of  conformal anomalies.

It has been shown \cite{edal} that the 
back-reaction for a classical bosonic string in $3+1$ dimensions has a
logarithmic divergence when the spacetime coordinate 
$x\rightarrow X(\sigma)$ that is, when we
approach the core of the string.\ This divergence is absorbed 
into a renormalization of the string tension.\ Copeland et al \cite{edal} showed
that by demanding that both 
it and the divergence in the energy-momentum tensor
vanish forces the string to have the couplings of compactified $N=1$,
$D=10$ supergravity.\ In this paper we are able to see that when we take into
account the quantum nature of the strings we lose all  
information regarding the position of the
string and therefore any divergences that may appear when one calculates
the back-reaction of quantum strings are not related to our
position with respect to the position of the string.

In the present work, the energy-momentum tensor of the string,
${\hat T}^{\mu\nu}$, is a quantum operator and it may be regarded 
as the vertex operator for the
emission (absorption) of gravitons in the case of closed strings or, for
the open string case, as the vertex operator for the
emission (absorption) of massive spin 2 particles.\ We compute 
its expectation value in one-particle string
states, choosing for the string center of mass wave function a wave packet
centered at the origin.  

Our results dramatically depend on the mass of the string state
chosen. For massive states the expectation value of the string
energy-momentum tensor identically vanishes at leading order (genus
zero), thus massive states of closed and open
strings give no contribution to ${\hat T}^{\mu\nu}$.

Massless string states (such as gravitons, photons  or dilatons)
yield  a non-zero results which turn out to be spin independent, that is the
same expression for  gravitons, photons  and dilatons.

We consider spherically symmetric and cylindrically symmetric configurations.
The components for the string
energy density and energy flux behave like massless waves, with the string 
energy being radiated outwards as a massless lump peaked at $r=t$
(for  the spherically symmetric case). We
provide integral representations for $<{\hat T}^{\mu\nu}(r,t)>$
[eq.(\ref{reprint})]. 
After exhibiting the tensor structure of  $<{\hat T}^{\mu\nu}>$
[eqs.(\ref{rotaT}) and (\ref{forfac})], the asymptotic behaviour of  
$<{\hat T}^{\mu\nu}(r,t)>$ 
for $ r \to \infty $ and $ t $ fixed and for  $ t \to \infty $ with $
r $ fixed is computed. In the first regime the energy density and the stress
tensor decay as $ r^{-1} $ whereas the energy flux decays as  $ r^{-2} $.
For  $ t \to \infty $ with $ r $ fixed, the energy density tends to $ 0^- $
as $ t^{-6} $. That is, the spherical wave leaves behind a rapidly
vanishing negative energy density.

For cylindrically symmetric configurations,  $<{\hat T}^{\mu\nu}(\rho,t)>$
progagates as outgoing (plus ingoing ) cylindrical waves. For large $
\rho $ and fixed $ t $ the energy density  decays
as $1/\rho$ and the energy flux decays as $1/\rho^2$. For large   $ t
\to \infty $ with $ \rho $ fixed,  the same phenomenom of a  negative
energy density vanishing as $ t^{-6} $ appears.

We restrict ourselves to bosonic strings in this paper. We expect
analogous results from superstrings, since only massless string
states contribute to the expectation values of  ${\hat T}^{\mu\nu}$. 

The structure of this paper is as follows: in section~2, we consider the
string energy-momentum tensor as a quantum operator and discuss its
quantum ordering problems, observing that
it can be regarded as a vertex operator.\ We calculate the expectation
value of the energy-momentum tensor for different physical  states of
the strings (for both open and closed strings). \ In
section~3, we calculate the string energy-momentum tensor expectation
value for massless states in spherically symmetric configurations, whilst
in section~4 we calculate it for cylindrically symmetric configurations
which are relevant when we study for example cosmic strings which are
essentially very long strings.\ Finally, in section~5 some final
remarks about our results are stressed.

\section{The string energy-momentum tensor.}

The energy-momentum tensor for a classical bosonic string with tension
$(\alpha')^{-1}$ is given by
\begin{equation}\label{six}
    T^{\mu\nu}(x)=\frac{1}{2\pi\alpha'}\int d\sigma d\tau   \;
(\dot{X}^{\mu}\dot{X}^{\nu}-X'^{\mu}X'^{\nu})\delta(x-X(\sigma,\tau))
\end{equation}
and the string coordinates are given in Minkowski space-time by
\begin{equation}\label{seven}
    X^{\mu}(\sigma,\tau)=q^{\mu}+2\alpha'p^{\mu}\tau
    +i\sqrt{\alpha'}\sum_{n\neq
    0}\frac{1}{n}[\alpha^{\mu}_{n}\;e^{-in(\tau-\sigma)}
    +\tilde{\alpha}^{\mu}_{n}\;e^{-in(\tau+\sigma)}]
\end{equation}
for closed strings and
\begin{equation}\label{sevena}
    X^{\mu}(\sigma,\tau)=q^{\mu}+2\alpha'p^{\mu}\tau+
    i\sqrt{\alpha'}\sum_{n\neq
    0}\frac{1}{n}\; \alpha^{\mu}_{n}\;e^{-in\tau}\cos n\sigma
\end{equation}
for open strings.
\vspace*{.5cm}
For closed strings we can set $\alpha'=1/2$, so 
inserting eq.~(\ref{seven}) in eq.~(\ref{six}) and rewriting the four 
dimensional delta function in integral form we obtain for closed strings,
\begin{eqnarray}
T^{\mu\nu}(x)=\frac{1}{\pi}\int d\sigma
d\tau\frac{d^{4}\lambda}{(2\pi)^{4}}
\{p^{\mu}p^{\nu}+\frac{p^{\mu}}{\sqrt{2}}\sum_{n\neq0}
[\alpha^{\nu}_{n}e^{-in(\tau-\sigma)}+\tilde{\alpha}^{\nu}_{n}
e^{-in(\tau+\sigma)}]+\nonumber\\
 +\sum_{n\neq0}(\alpha^{\mu}_{n}
e^{-in(\tau-\sigma)}+\tilde{\alpha}^{\mu}_{n}
e^{-in(\tau+\sigma)})\frac{p^{\nu}}{\sqrt{2}}
+\nonumber\\+\sum_{n\neq0}\sum_{m\neq0}[\alpha^{\mu}_{n}
\tilde{\alpha}^{\nu}_{m}e^{-in(\tau-\sigma)}e^{-im(\tau+\sigma)}
+\tilde{\alpha}^{\mu}_{n}\alpha^{\nu}_{m}e^{-in(\tau+\sigma)}
e^{-im(\tau-\sigma)}]\}\;e^{i\lambda\cdot x}\;e^{-i\lambda\cdot
X(\sigma,\tau)}.\label{twelve}
\end{eqnarray}
We can write
$$X(\sigma,\tau)=X_{cm}+X_{+}+X_{-},$$
where $X_{cm}=q+p\tau$ is the centre of mass coordinate and 
$X_{+}$ and $X_{-}$ refer to the terms with $\alpha_{n>0}$ and
$\alpha_{n<0}$ in $X(\sigma,\tau)$ respectively.
\ In this way, we can 
see now that our energy-momentum tensor has the same form as that of a vertex
operator, when we recall that the vertex operator for closed strings 
has to have conformal dimension 2 whereas for open strings it has to 
have conformal dimension 1 [10] in order that the string is anomaly free.

Eq.(\ref{six}) is meaningful at the classical level.\ However, at 
the quantum level, one must be careful with the order of the operators 
since $\dot{X}^{\mu}$ and $\dot{X}^{\nu}$
do not commute with $X(\sigma,\tau)$. We shall define the quantum
operator $ {\hat T}^{\mu\nu}(x) $ by symmetric ordering. That is,
 \begin{eqnarray}\label{Tcua}
   {\hat T}^{\mu\nu}(x)&\equiv& \frac{1}{2\pi\alpha'}\int d\sigma d\tau \cr 
\left\{{1 \over 3}\left[ 
\dot{X}^{\mu}\dot{X}^{\nu}\delta(x-X(\sigma,\tau))
\right. \right.&+&\left.\left. 
\dot{X}^{\mu}\delta(x-X(\sigma,\tau))\dot{X}^{\nu}+
\delta(x-X(\sigma,\tau))\dot{X}^{\mu}\dot{X}^{\nu}\right] \right. \cr
&-&\left.\begin{array}{l}
X'^{\mu}X'^{\nu}\delta(x-X(\sigma,\tau))\\
\end{array}\right\}
\end{eqnarray}
This definition ensures hermiticity:
$$
{\hat T}^{\mu\nu}(x)^{\dag}= {\hat T}^{\mu\nu}(x)
$$
Let us consider a string on a mass and spin eigenstate with a center
of mass wave function
$\varphi(\vec{p})\delta(p^{0}-\sqrt{\vec{p}^{2}+m^{2}}) $.\ An 
on-shell scalar string state is then 
$$
 | \Psi\rangle = \int d^4 p \;
\varphi(\vec{p})\; \delta(p^{0}-\sqrt{\vec{p}^{2}+m^{2}})\;  | \vec{p} \rangle
.$$
Here we assume the extra space-time dimensions (beyond four) to be
appropriately compactified, and consider string states in the physical
(uncompactified) four dimensional Minkowski space-time.
 
Now, normal ordering eq.(\ref{twelve}) using eq.(\ref{seven}) and 
taking the expectation value with respect to the fundamental scalar state 
(tachyonic), we get 
\begin{eqnarray}
\frac{\langle \Psi | {\hat T}^{\mu\nu}(x) | \Psi \rangle}
{\langle\Psi |\Psi\rangle} \equiv \langle
{\hat T}^{\mu\nu}(x)\rangle 
=\frac{1}{3\pi}\int\frac{d^{4}\lambda}{(2\pi)^{4}}\; d^4p_{1}d^4p_{2}\;d\sigma 
d\tau \; e^{i\lambda\cdot
x}\;[p^{\mu}_{1}p^{\nu}_{1}+p^{\mu}_{2}p^{\nu}_{2} +
p^{\mu}_{1}p^{\nu}_{2}]\nonumber\\  
\langle p_{1}|e^{-i\lambda\cdot X_{cm}}|p_{2}\rangle\;
\varphi^{*}(\vec{p_{1}})\varphi(\vec{p_{2}})
\; \delta(p^{0}_{1}-\sqrt{\vec{p_{1}}^{2}+m^{2}_{1}})
\delta(p^{0}_{2}-\sqrt{\vec{p_{2}}^{2}+m^{2}_{2}}).
\label{fifteen} 
\end{eqnarray} 
Writing
$$
\langle p_{1}|e^{-i\lambda\cdot X_{cm}}|p_{2}\rangle =
e^{-i\frac{\tau\lambda^{2}}{2}-i\lambda p_{2}\tau}
\delta^{4}(\lambda+p_{1}-p_{2}),
$$
eq.(\ref{fifteen}) becomes

\begin{eqnarray}
\langle {\hat T}^{\mu\nu}(x)\rangle =\frac{2}{3}\int\frac{d^{4}\lambda}
{(2\pi)^{4}}\; d^4p_{1}d^4p_{2}\;d\tau \; e^{i\lambda\cdot x}\; 
[p^{\mu}_{1}p^{\nu}_{1}+p^{\mu}_{2}p^{\nu}_{2} +
p^{\mu}_{1}p^{\nu}_{2}]\nonumber\\  
e^{-i\frac{\tau\lambda^{2}}{2}-i\lambda p_{2}\tau}
\; \delta^{4}(\lambda+p_{1}-p_{2}) 
\varphi^{*}(\vec{p_{1}})\varphi(\vec{p_{2}})\; 
\delta(p^{0}_{1}-\sqrt{\vec{p_{1}}^{2}+m^{2}_{1}})
\delta(p^{0}_{2}-\sqrt{\vec{p_{2}}^{2}+m^{2}_{2}})\; .\nonumber
\end{eqnarray}
Performing the $\lambda$ and $\tau$ integrals  
\begin{eqnarray}
\langle {\hat T}^{\mu\nu}(x)\rangle =
\frac{4}{3(2\pi)^{3}}\int d^4p_{1}d^4p_{2}\; e^{i(p_{2}-p_{1})\cdot x}
\; [p^{\mu}_{1}p^{\nu}_{1}+p^{\mu}_{2}p^{\nu}_{2} +
p^{\mu}_{1}p^{\nu}_{2}]\nonumber\\ \delta((3p_{2}-p_{1})\cdot(p_{2}-p_{1}))
\;\varphi^{*}(\vec{p_{1}})\varphi(\vec{p_{2}})\; 
\delta(p^{0}_{1}-\sqrt{\vec{p_{1}}^{2}+m^{2}_{1}})
\delta(p^{0}_{2}-\sqrt{\vec{p_{2}}^{2}+m^{2}_{2}}).
\label{nineteena}
\end{eqnarray} 
The calculation for open strings is obtained by substituing
eq.(\ref{sevena}) into eq.(\ref{Tcua}), with the result (setting $\alpha'=1$)

\begin{eqnarray}
\langle {\hat T}^{\mu\nu}(x)\rangle =
\frac{2}{3(2\pi)^{3}}\int d^4p_{1}d^4p_{2}\; e^{i(p_{2}-p_{1})\cdot x}\; 
[p^{\mu}_{1}p^{\nu}_{1}+p^{\mu}_{2}p^{\nu}_{2} +
p^{\mu}_{1}p^{\nu}_{2}]\nonumber\\ \delta((3p_{2}-p_{1})\cdot(p_{2}-p_{1}))
\;\varphi^{*}(\vec{p_{1}})\varphi(\vec{p_{2}})\; 
\delta(p^{0}_{1}-\sqrt{\vec{p_{1}}^{2}+m^{2}_{1}})
\delta(p^{0}_{2}-\sqrt{\vec{p_{2}}^{2}+m^{2}_{2}})\; .\nonumber
\end{eqnarray}
The extra factor two in eq.(\ref{nineteena}) comes from the fact that,
for closed strings, we have two independent sets of oscillation modes.

Now, in general, we want to take the expectation value with respect to   
particle  states with higher (mass)$^2$ 
and spin than the tachyonic case.\ Since we are interested
in expectation values we must have the same particle in both states,
hence $m_{1}=m_{2}=m$.\ As we shall show later, only 
for massless particle states has the energy-momentum tensor a
non-zero expectation value. Therefore, let us first 
consider massless string states.

 For the closed string there is the graviton
  \begin{equation}\label{graviton}
|\vec{p} ; s \rangle = P^{i l}_s (n)\;
\tilde{\alpha}^l_{-1}\;\alpha^i_{-1}\;|\vec{p}\rangle 
\end{equation}
and the dilaton
 \begin{equation}\label{dilaton}
|\vec{p}  \rangle = P^{i l}(n)\;
\tilde{\alpha}^l_{-1}\;\alpha^i_{-1}\;|\vec{p}\rangle 
\end{equation}
where $\vec{p}$ is the momentum ($p^2 = 0$), $s= \pm$ labels the
graviton helicity, 
$ P^{i l}_s (n), \; 1\leq i, l \leq 3$ projects into the spin 2 
graviton states and 
$ P^{i l}(n)$ into the (scalar) dilaton state,
\begin{eqnarray}
n^i &\equiv& {{p^i} \over {|\vec{p}|}} \cr \cr
 P^{i l}_s (n) &=&  P^{ l i}_s (n) \quad , \quad  n^i\, P^{i l}_s (n)= 0 \cr\cr
 P^{l l}_s (n) &=& 0  \quad , \quad  
P^{i l}_s (n) P^{i l}_{s'} (n) = \delta_{s s'} \cr \cr
 P^{i l}_s (n) &=&  P^{i l}_s (-n)  \quad , \quad 
 P^{i l}(n) \equiv  \delta^{i l} - n^i n^l \; . \nonumber
\end{eqnarray}
The massless vector states (photons) for open strings are given by
$$
|\vec{p} ; i \rangle = P^{i l} (n) \; \alpha^l_{-1}\,|\vec{p}\rangle \; .
$$
In analogy to the tachyon case eq.(\ref{fifteen}) we obtain for closed strings
\begin{eqnarray}
\langle  \vec{p_1} | \tilde{\alpha}^l_{1}\alpha^i_{1}
{\hat T}^{\mu\nu}(x) \tilde{\alpha}^j_{-1}\alpha^m_{-1}
|\vec{p_2} \rangle
=\frac{1}
{12\pi}\int\frac{d^{4}\lambda}{(2\pi)^{4}}d^4p_{1}d^4p_{2}\; d\sigma 
d\tau \;A^{ljim}\;
[p^{\mu}_{1}p^{\nu}_{1}+p^{\mu}_{2}p^{\nu}_{2} +
p^{\mu}_{1}p^{\nu}_{2}]\nonumber\\ \langle p_{1}|
e^{i\lambda\cdot X}e^{-i\lambda\cdot X_{cm}}|p_{2}\rangle\;
\varphi^{*}(\vec{p_{1}})\;\varphi(\vec{p_{2}}) \; 
\delta(p^{0}_{1}-\sqrt{\vec{p_{1}}^{2}+m^{2}})\;
\delta(p^{0}_{2}-\sqrt{\vec{p_{2}}^{2}+m^{2}}) \; ,
\label{antproy}
\end{eqnarray}
where
$$
A^{ljim}=4\delta^{lj}\delta^{im}-2\delta^{lj}\lambda^{i}\lambda^{m}
-2\delta^{im}\lambda^{l}\lambda^{j} +
\lambda^{l}\lambda^{j}
\lambda^{i}\lambda^{m} \; .
$$
Whereas for the open string case we get, 
 
\begin{eqnarray}
\langle \vec{p_{1}}|\alpha^i_{1} {\hat T}^{\mu\nu}(x)\alpha^m_{-1}
| \vec{p_{2}} \rangle=\frac{2}{3}
\int d^4p_{1}d^4p_{2} \; \frac{e^{i(p_{2}-p_{1})\cdot
X}}{(2\pi)^{3}} \; [\delta^{i m}\; (p^{\mu}_{1}p^{\nu}_{1}
+p^{\mu}_{2}p^{\nu}_{2} +p^{\mu}_{1}p^{\nu}_{2})\nonumber\\ -\frac{3}{2}
\eta^{\mu\nu}(p_{2}-p_{1})^{i}(p_{2}-p_{1})^{m}] \; 
\delta((3p_{2}-p_{1})\cdot
(p_{2}-p_{1}))\nonumber\\ 
\varphi^{*}(\vec{p_{1}})\varphi(\vec{p_{2}}) \; 
\delta(p^{0}_{1}-\sqrt{\vec{p_{1}}^{2}+m^{2}})
\delta(p^{0}_{2}-\sqrt{\vec{p_{2}}^{2}+m^{2}}).
\label{unop}
\end{eqnarray}
having performed the $\sigma$, $\tau$ and $\lambda$ integrations.

Projecting on the massless physical states 
(\ref{graviton}) and (\ref{dilaton}) and
integrating over $p ^{0}_{1}$ and 
$p ^{0}_{2}$, eqs.(\ref{antproy}) and (\ref{unop}) become 
\begin{eqnarray}
\langle p_1, s |
{\hat T}^{\mu\nu}(x)| p_2 , s \rangle &=&
\frac{1}{6(2\pi)^{3}}\int d^{3}p_{1}d^{3}p_{2}e^{i(p_{2}-p_{1})\cdot
x}\;  
(p^{\mu}_{1}p^{\nu}_{1}+p^{\mu}_{2}p^{\nu}_{2} +
p^{\mu}_{1}p^{\nu}_{2})\nonumber\\ 
& & \varphi^{*}(\vec{p_{1}})\varphi(\vec{p_{2}})\; \delta(p_{2}\cdot p_{1}),
\label{dosa}
\end{eqnarray}
\begin{eqnarray}
\langle p_1| {\hat T}^{\mu\nu}(x)| p_2\rangle&=&{1\o {6 (2\pi)^3}}\; \int
d^3p_{1} d^3p_{2} 
e^{i(p_{2}-p_{1})\cdot x}\; 
(p^{\mu}_{1}p^{\nu}_{1}+p^{\mu}_{2}p^{\nu}_{2} +
p^{\mu}_{1}p^{\nu}_{2})\nonumber\\  
& & \varphi^{*}(\vec{p_{1}})\varphi(\vec{p_{2}})\; \delta(p_{2}\cdot p_{1})
\label{dos}
\end{eqnarray}
respectively.\ It is easy to check that this energy-momentum 
tensor is indeed conserved.

We will evaluate eq.(\ref{dosa}) and eq.(\ref{dos}) shortly.\ For the
massive case ($m\neq 0$) we will now show that the equivalent result has to be
identically zero.\ It is convenient to 
parametrize the momenta $  p_1 $ and  $  p_2 $ for $m \neq 0$ in
eq.(\ref{antproy}) and eq.(\ref{unop}) as follows:
$$
p_1 = m (\cosh u, \sinh u \, {\hat u_1}) \quad , \quad
p_2 = m (\cosh v, \sinh v \, {\hat u_2})
$$
with $ {\hat u_1}$ and  $ {\hat u_2}$  unit three-dimensional
vectors       and $u,v\geq 0$. Then,
$$
p_{2}\cdot p_{1}-m^2={{m^2}\o 2} \left[(1 -  {\hat u_1}\cdot {\hat u_2})
\cosh(u+v) +(1 +  {\hat u_1}\cdot {\hat u_2}) \cosh(u-v) - 2 \right]
$$
The only real root for   $m \neq 0$ corresponds to
$$
  {\hat u_1} =  {\hat u_2} \quad ,  \quad u - v = 0
$$
and arbitrary $u + v$. This means that the delta function in eq.(\ref{dos}) 
{\bf sets} $p_1 =p_2 $ and we arrive to a constant ($x$-independent) result
for $\langle {\hat T}^{\mu\nu}(x)\rangle$. 

More precisely, in spherical coordinates
$$
  {\hat u_1} =(\cos\alpha \sin\gamma, \sin\alpha \sin\gamma, \cos \gamma)
\quad , \quad
  {\hat u_2} =(\cos\beta \sin\delta, \sin\beta \sin\delta, \cos \delta)
$$
and we find
$$
\delta(p_{2}\cdot p_{1}-m^2)= {{8\sqrt{2}\pi}\o{\sin\gamma(\cosh t-1)m^3}}\,
\sqrt{|p_{2}\cdot p_{1}-m^2|}\;\delta(\alpha-\beta)\delta(\gamma-\delta)
\delta(u-v)
$$
So, actually the expectation value  $\langle {\hat T}^{\mu\nu}(x)\rangle$
 vanishes for $ m^2 \neq 0 $
since the argument of the $\sqrt{}$ vanishes at
$\alpha=\beta, \gamma =\delta, u = v $.

Actually, the only reasonable constant ($x$-independent) value for $\langle 
{\hat T}^{\mu\nu}(x)\rangle$ describing a localized object is precisely zero.\
{F}rom this result we can see that open  as well as the closed string
massive states do not contribute to the expectation value of $ {\hat
T}^{\mu\nu}(x)$.

It seems that any particle emitted by a freely moving quantum string
is  massless  such as photons, gravitons and dilatons.
 
\section{${\hat T}^{\mu\nu}(x)$ for massless string states  
in spherically symmetric configurations.}

Let us now consider the expectation value of the energy-momentum tensor 
for the massless closed string state given by eq.(\ref{dosa}) and
eq.(\ref{dos}).\  
Notice that such  expectation value is independent of the value of the
particle spin (zero, one or two).

For such a case, it is convenient to use the parametrization:
$$
p_1 = E_1 (1,  {\hat u_1}) \quad , \quad
p_2 = E_2 (1, {\hat u_2})
$$
with $ E_1 ,  E_2 \geq 0$.

Then we see that $ p_{2}\cdot p_{1} = 0 $ implies 
$  {\hat u_1} \cdot  {\hat u_2} = 1$ (unless   $ E_1 $ or  $ E_2 $
vanishes). Therefore   $ E_1 $ {\bf does not need} to be equal to  $ E_2 $ and
we find here a non-constant result.

More precisely, we find in spherical coordinates
$$
\delta(p_{2}\cdot p_{1})={{2\pi}\o{E_1 E_2 \sin \gamma}} \;
\delta(\alpha-\beta)\delta(\gamma-\delta)
$$
Inserting this result in eq.(\ref{dosa}) and integrating over $ {\hat u_2} $
yields
\begin{eqnarray}\label{reprint}
\langle {\hat T}^{\mu\nu}(x)\rangle &=& 
\frac{\pi }{3(2\pi)^{3}} \int_0^{\infty} E_1 dE_1 
\int_0^{\infty} E_2 dE_2 \;
\varphi^{*}(E_1,{\hat u_{1}})\; \varphi(E_2,{\hat u_{1}})\;
e^{i(E_2-E_1)(t-\vec{x}\cdot{\hat u_{1}})} \nonumber\\ 
& & \left[p^{\mu}_{1}p^{\nu}_{1}
+p^{\mu}_{2}p^{\nu}_{2}+p^{\mu}_{1}p^{\nu}_{2}\right]
\;d{\hat u_{1}}
\label{tmassles}
\end{eqnarray}
where $d{\hat u_{1}} \equiv \sin\gamma \, d\gamma \, d\alpha , \;p_1 =
E_1 (1,  {\hat u_1}) $ and now $p_2 = E_2 (1, {\hat u_1})$.

\hspace*{-.6cm}Let us consider for simplicity 
spherically symmetric wave packets
$ \varphi(E,{\hat u}) = \varphi(E)$.\ If we take $\vec{x}=(0,0,r)$ 
we can then integrate over the angles in eq.(\ref{tmassles}) with the result
\begin{eqnarray}\label{esfsim0}
 \langle {\hat T}^{00}(t,r)\rangle&=& \frac{1}{6\pi}  
\int_0^{\infty} E_1 \, dE_1 \int_0^{\infty} E_2 \, dE_2 \;
\varphi^{*}(E_1)\varphi(E_2)\; e^{i(E_2-E_1)t} \nonumber\\  
& & {{\sin(E_2-E_1)r}\o{(E_2-E_1)r}} \;
\left[ E^{2}_{1} + E^{2}_{2} +  E_1 E_2 \right]
\end{eqnarray}
We can relate the result for
arbitrary $x=(t, {\vec x})$ with the special case   $ x = (t,0,0,z) $
using rotational invariance as follows,
\begin{eqnarray}\label{rotaT}
\langle {\hat T}^{0 i}(x)\rangle &=& {\hat x}^i \; C(t,r) \quad ,
\quad i=1,2,3 ,\cr \cr
\langle {\hat T}^{i j}(x)\rangle &=& \delta^{ij} \; A(t,r) +  {\hat
x}^i \, {\hat x}^j  \; B(t,r) \; \; , \; i,j = 1,2,3.
\end{eqnarray}
Here
\begin{eqnarray}\label{defABC}
 C(t,r)&=& \langle {\hat T}^{03}(t,r=z)\rangle \cr \cr
 A(t,r)=  \langle {\hat T}^{22}(t,r=z)\rangle \quad &,& \quad
 B(t,r)= \langle {\hat T}^{33}(t,r=z)\rangle -  \langle {\hat
T}^{11}(t,r=z)\rangle
\end{eqnarray}
with ${\hat x}^{i}=\frac{x^i}{r}$ the unit vector, and 
\begin{eqnarray}\label{esfsimr}
\langle {\hat T}^{11}(t,r=z)\rangle&=& -  \frac{1}{6\pi}  \;
\int_0^{\infty} E_1 \, dE_1 \int_0^{\infty} E_2 \, dE_2 \;
\varphi^{*}(E_1)\varphi(E_2)\;
{{e^{i(E_2-E_1)t}}\o{(E_2-E_1)^{2}r^{2}}} \nonumber\\ 
& & \left[\cos(E_2-E_1)r - {{\sin(E_2-E_1)r}\o{(E_2-E_1)r}}\right]
 \left[  E^{2}_{1} + E^{2}_{2} + E_1 E_2 \right], \cr
\langle {\hat T}^{33}(t,r=z)\rangle&=&  \frac{1}{6\pi}\,  
\int_0^{\infty}  E_1 \,dE_1 \int_0^{\infty}  E_2 \,dE_2 \;
 \left[  E^{2}_{1} + E^{2}_{2} + E_1 E_2 \right]\nonumber\\
& &  \varphi^{*}(E_1)\varphi(E_2)\;
{{e^{i(E_2-E_1)t}}\o{(E_2-E_1)r}} \;
\left[ \sin(E_2-E_1)r+2{{\cos(E_2-E_1)r}\o{(E_2-E_1)r}} \right.\nonumber\\ 
& & \left. - 2{{\sin(E_2-E_1)r}\o{(E_2-E_1)^{2}r^{2}}}\right], \nonumber\\
 \langle {\hat T}^{03}(t,r=z)\rangle&=&  \frac{i}{6\pi} \,  
\int_0^{\infty}  E_1 \, dE_1 \int_0^{\infty} E_2 \, dE_2  \; \left[ 
E^{2}_{1} + E^{2}_{2} +  E_1 E_2 \right] \nonumber\\
& &  \varphi^{*}(E_1)\varphi(E_2)\;
{{e^{i(E_2-E_1)t}}\o{(E_2-E_1)r}} \;
\left[\cos(E_2-E_1)r -
{{\sin(E_2-E_1)r}\o{(E_2-E_1)r}}\right].
\end{eqnarray}
The other components satisfy 
$\langle {\hat T}^{22}(t,r=z)\rangle=\langle {\hat T}^{11}(t,r=z)\rangle$,
 $\langle {\hat T}^{01}(t,r=z)\rangle=\langle {\hat T}^{02}(t,r=z)\rangle=
\langle {\hat T}^{12}(t,r=z)\rangle=
\langle {\hat T}^{13}(t,r=z)\rangle=
\langle {\hat T}^{23}(t,r=z)\rangle=0$, as they must be from rotational
invariance.

\ As we can see the trace of the expectation value of the string
energy-momentum 
tensor vanishes. In other words the trace 
of the energy-momentum tensor  induced by 
a quantum string vanishes. \ Only massless particles
are responsible for the  field created by the string.

Notice that 
$$
3 A(t,r) + B(t,r) =  \langle {\hat T}^{00}(t,r)\rangle 
$$ 
due to the tracelessness of the energy-momentum tensor. 

The $r$ and $t$ dependence in the invariant functions $  \langle {\hat
T}^{00}(t,r)\rangle,  A(t,r),\;
B(t,r)$ and $C(t,r)$ writing eqs.(\ref{esfsim0}), (\ref{defABC}) and
(\ref{esfsimr}) as 

\begin{eqnarray}\label{forfac}
\langle {\hat T}^{00}(t,r)\rangle&=&{1\o r}\; \left[ F(t+r) -
F(t-r)\right],\cr\cr 
A(t,r) &=&        -{1\o{ r^2}} \left[ H(t+r) + H(t-r)\right] 
        +{1\o{ r^3}} \left[ E(t+r) - E(t-r)\right],\cr\cr
B(t,r)&=&{1\o r}\; \left[ F(t+r) - F(t-r)\right]
         +{3\o{ r^2}} \left[ H(t+r) + H(t-r)\right]
 -{3\o{ r^3}} \left[ E(t+r) - E(t-r)\right],\cr\cr
C(t,r)&=&-{1\o r} \left[ F(t+r) + F(t-r)\right]-
{1\o{ r^2}} \left[ H(t+r) - H(t-r)\right]\; .  
\end{eqnarray}

where

\begin{eqnarray}\label{fint}
F(x) &=& { 1 \over{12 \pi}}\; 
\int_0^{\infty} E_1 \, dE_1 \int_0^{\infty} E_2 \, dE_2 \;
\varphi(E_1)\varphi(E_2)\nonumber\\  
& & {{\sin(E_2-E_1)x}\o{(E_2-E_1)}} \;
\left[ E^{2}_{1} + E^{2}_{2} +  E_1 E_2 \right]\; .
\end{eqnarray}

Notice that $ F(x) = - H'(x) $ and $ H(x) = E'(x) $. These relations
guarantee the conservation of $ \langle{\hat T}^{\mu\nu}(x)\rangle $,
$$
{{\partial}\over{\partial t}} \langle{\hat T}^{0\nu}(x)\rangle+
 {{\partial}\over{\partial x^i}}\langle{\hat T}^{i\nu}(x)\rangle=0\; .
$$

We choose a  real wave-packet $\varphi(E)$ decreasing fast with $E$ and
typically peaked at $E=0$. 
For example a gaussian wave-packet $\varphi(E)$:

\begin{equation}\label{pagau}
\varphi(E) = \left( {{2\alpha}\o{\pi}} \right)^{3/4}\;e^{-\alpha E^2}, 
\end{equation}

We see from eq.(\ref{forfac}) that the energy density 
$\langle {\hat T}^{00}(r,t)\rangle$ and the energy flux $\langle
{\hat T}^{0i}(r,t)\rangle$ behave like spherical waves describing 
the way a string massless
state spreads out starting from the initial wave-packet we choose. 

In order to compute the asymptotic behaviour of the function $F(x)$ 
we change in eq.(\ref{fint}) the integration variables 
$$
E_2-E_1 = v \tau / x \quad , \quad E_2+E_1 = v\; .
$$
We find
\begin{eqnarray}
F(x) &=& { 1 \over{192 \pi}}\; 
\int_0^{\infty} v^4 \, dv \; \int_0^{x} {{d\tau}\over \tau}\;
\varphi(\frac{v}{2}[1
+\frac{\tau}{x}])\varphi(\frac{v}{2}[1
-\frac{\tau}{x}])\nonumber\\   
& & \sin(v\tau) \;
\left[3 - 2 {{\tau^2}\over {x^2}} -  {{\tau^4}\over {x^4}}\right]\; .
\nonumber
\end{eqnarray}
Now, we can let $x \to \infty$ with the result
$$
F(x)  \buildrel{x\to \pm\infty}\over = \pm \frac14\; \int_0^{\infty} E^4 \,
dE\; \varphi(E)^2 + {{ 2\, \varphi(0)^2}\over{15 \pi \; x^5}} + O({1
\over {x^7}}) 
$$

We find through similar calculations,
\begin{eqnarray}
H(x)  \buildrel{x\to \pm\infty}\over = -\frac{M}4\;|x| + \frac{N}{4\pi}
+  {{ \, \varphi(0)^2}\over{30 \pi \; x^4}} + O({1 \over {x^6}}) \; ,
\cr \cr
E(x)  \buildrel{x\to \pm\infty}\over = -\frac{M}{8}\;x^2\; sign(x) +
\frac{N\; x}{4\pi} 
-  {{ \, \varphi(0)^2}\over{90 \pi \; x^3}} + O({1 \over {x^5}})  \; .
\nonumber
\end{eqnarray}

Here,
$$ 
M\equiv \int_0^{\infty} E^4 \,dE\; \varphi(E)^2\; \quad \mbox{and}
\quad N \equiv \int_0^{\infty} E^3 \,dE\; \varphi(E)^2\; .
$$
For the gaussian wave  packet (\ref{pagau}) they take the values
$$
M = {3 \o {16 \pi \alpha}} \quad \mbox{and} \quad N = {1 \o {
(2\pi)^{3/2}\; \sqrt{\alpha}}} \; .
$$

To gain an insight into the behaviour of 
$\langle\hat{T}^{\mu\nu}\rangle$, we consider
the limiting cases: $ r\rightarrow\infty , \;  t $ fixed,
 $ r = t \rightarrow\infty $ 
and $ t\rightarrow\infty , \;  r $ fixed
with the following results:

a) $ r\rightarrow\infty , \; t$ fixed 

\begin{eqnarray}
\langle {\hat T}^{00}(r,t)\rangle &\buildrel{r\to \infty}\over =& 
 \frac{M}{2\;r}+  {{ 4 \; \varphi(0)^2}\over{15 \pi \; r^6}} +
O({1\over {r^7}})  \quad , \cr \cr
A(r,t) &\buildrel{r\to \infty}\over =&
\frac{M}{4\; r}\left(1 - {{t^2} \o {r^2}} \right)- {{ 4\;
\varphi(0)^2}\over{45 \pi \; r^6}} + O({1 \over {r^7}})  \quad , \cr \cr
B(r,t)&\buildrel{r\to \infty}\over=&
-\frac{M}{4\; r} \left(1 - {{3\; t^2} \o {r^2}} \right)+ {{ 8\;
\varphi(0)^2}\over{15 \pi \; r^6}} + O({1 \over {r^7}})   \nonumber
\end{eqnarray}
and 
$$
C(r,t) \buildrel{r\to \infty}\over=
\frac{M \; t}{2\; r^2} + O({1\over {r^7}}) \quad .
$$

b) $r=t$ and large

\begin{eqnarray}
\langle {\hat T}^{00}(r,t)\rangle &\buildrel{r=t\to \infty}\over=&
\frac{M}{4\; r} +  {{ \varphi(0)^2}\over{240 \pi \; r^6}} + O({1 \over
{r^7}})  \cr \cr
A(r,t)&\buildrel{r=t\to \infty}\over =&  - {{ 5 \;
\varphi(0)^2}\over{1440 \pi \; r^6}} + O({1 \over 
{r^7}})  \cr \cr
B(r,t)&\buildrel{r=t\to \infty}\over =&  \frac{M}{4\; r} +  {{ 7 \; 
\varphi(0)^2}\over{480 \pi \; r^6}} + O({1 \over 
{r^7}})  \cr \cr
C(r,t)&\buildrel{r=t\to \infty}\over =&
 \frac{3 M}{4\; r} + {N \o{4\pi\; r^2}}-   {{ \varphi(0)^2}\over{480
\pi \; r^6}}  + O({1 \over{r^7}}) \nonumber
\end{eqnarray}

c) $t\rightarrow\infty, r$ fixed

\begin{eqnarray}
\langle {\hat T}^{00}(r,t)\rangle &\buildrel{t\to \infty}\over=&
- \frac{4 \varphi(0)^2}{3\pi}\; {1 \o {t^6}} + O(t^{-8})\cr \cr
A(r,t)&\buildrel{t\to \infty}\over=& 0 + O(t^{-6}) \quad , \cr \cr
B(r,t)&\buildrel{t\to \infty}\over=& 0  + O(t^{-6})\quad ,\cr \cr
C(r,t)&\buildrel{t\to \infty}\over=& 0  + O(t^{-7}) \quad . \nonumber
\end{eqnarray}

Thus, as we mentioned earlier the  energy density, the energy flux  and the
components of the stress tensor propagate 
as spherical outgoing waves. For  $t$ fixed, the  energy density
decays as $ r^{-1} $ while the  energy flux decays as  $ r^{-2} $.
This corresponds to a $r$-independent radiated energy for large $r$.

For $r$ fixed and large $t$, the energy density decays rapidly as
$O(1/t^6)$. We curiously find a  negative energy density in this
regime. The spherical wave seems to leave behind a small but negative
energy density. Notice that $ T^{00} $ is {\bf not} a positive definite
quantity for strings [see eq.(\ref{six})].\

It must be stressed that the expectation value of the energy-momentum
tensor is {\bf everywhere} finite. Even at the potentially troublesome 
limit,  for $r$ and $t$  approaching zero, a careful 
examination of eqs. (\ref{esfsim0}-\ref{esfsimr}) shows that {\bf all}  
$\langle {\hat T}^{\mu\nu}(x)\rangle$ components are finite constants in
such a limit.

These last results were derived using a gaussian shape for the wave
packet.\ It is clear that the results will be qualitatively similar
for any fast-decaying wave function $\varphi(E)$.

It should be noted that analogous but not identical results follow for
massless waves.\ A spherically symmetric solution of the wave
equation in $D=3+1$ dimensions 
$$
\partial^2 \phi(r,t) = 0
$$
takes the form,
\begin{equation}\label{solfi}
\phi(r,t) = {1 \over r} \left[ f(t-r) + g(t+r) \right]
\end{equation}
where $f(x)$ and $g(x)$ are arbitrary functions.\  
The energy-momentum tensor for such a massless scalar field can be
written as:
\begin{equation}\label{tmnfi}
 {\hat T}^{\mu\nu}(x)_{\phi}= \partial^{\mu}\phi \partial^{\nu}\phi-
{{\eta^{\mu\nu}}\over 2} (\partial\phi)^2.
\end{equation}
Inserting eq.(\ref{solfi}) into  eq.(\ref{tmnfi}) yields,
\begin{eqnarray}
 {\hat T}^{00}(r,t)_{\phi} &=& {1 \over {r^2}}\left\{ f'^2 +g'^2 + {{f+g}\over
{2r}}\left[ 2(f'-g')+ {{f+g}\over r} \right] \right\}. \cr \cr 
 {\hat T}^{0i}(r,t)_{\phi} &=& {{x^i}\over {r^3}}\;\left[ g'^2 -  f'^2
- \frac1r(f+g)(f' - g') \right]\cr \cr
 {\hat T}^{ij}(r,t)_{\phi} &=& \frac1{2r^2}\;\left[ \delta^{ij} +
{{2x^ix^j}\over {r^2}} \right]\;\left[f'-g'+\frac1r(f+g)\right]^2
\nonumber
\end{eqnarray}
We see that the string $ {\hat T}^{\mu\nu}$ scales as $1/r$ [eq.(\ref{forfac})]
whereas the field $ {\hat T}^{\mu\nu}_{\phi}$ scales as $1/r^2$.
Perhaps the slower $ {\hat T}^{\mu\nu}$ decay for strings can be related to the
fact that they are extended objects.

\section{${\hat T}^{\mu\nu}$ for massless string states in 
cylindrically symmetric configurations.}
Cosmic strings can be considered as essentially very long straight
strings in
(almost) cylindrically symmetric configurations.\ Although they
behave as fundamental strings only classically and in the
Nambu approximation (that is, zero string thickness), it
is important to study the expectation value of  $ {\hat T}^{\mu\nu}(x)$
for a cylindrically symmetric configuration.\ It would be interesting
to see if there is an equivalent quantum version of the deficit angle
found for cosmic strings \cite{shellard}.\ Consider
a cylindrically symmetric wave packet

\begin{equation}\label{paqcil}
\varphi(E,{\hat u}) = \varphi(E , \gamma).
\end{equation}
We can rewrite the integral (\ref{tmassles}) in a more convenient way to
analyse $\langle {\hat T}^{\mu\nu}(x)\rangle$ in a cylindrical
configuration as follows:

\begin{eqnarray}
\langle {\hat T}^{\mu\nu}(x)\rangle &=& 
\frac{1 }{24 \pi^2} \int_0^{\infty} E_1 dE_1 
\int_{0}^{\infty}E_2 dE_2\int_0^{\pi}
d\gamma\sin\gamma\int_0^{2\pi}d\alpha 
\;\varphi^{*}(E_1,\gamma)\; \varphi(E_2,\gamma)\nonumber\\
& & e^{i(E_2-E_1)t}\;e^{i(E_2-E_1)\rho\cos\alpha\sin\gamma} 
\left[p^{\mu}_{1}p^{\nu}_{1}
+p^{\mu}_{2}p^{\nu}_{2}+p^{\mu}_{1}p^{\nu}_{2}\right]
\label{tmassles2}
\end{eqnarray}
where we have choosen ${\vec x} = (\rho,0,0)$ , ${\hat u}_1.{\vec x} = \rho
\cos\alpha\,\sin\gamma$, and $L$ is the radius of a large sphere.

Integrating over $ \alpha $ in eq.(\ref{tmassles2}) we find

\begin{eqnarray}\label{cilsim}
\langle {\hat T}^{00}(t,\rho)\rangle &=& \frac{1}{12\pi}  
\int_0^{\infty}E_1 \,  dE_1 \int_0^{\infty} E_2 \,
dE_2\int_0^{\pi}d\gamma\sin\gamma\;
\varphi^{*}(E_1,\gamma)\varphi(E_2,\gamma) \nonumber\\
& & \left[  E^{2}_{1} + E^{2}_{2} +  E_1 E_2 \right]
 J_0([E_1-E_2]\rho\,\sin\gamma)\; e^{i(E_2-E_1)t} \; ,\cr \cr
\langle {\hat T}^{33}(t,\rho)\rangle &=& \frac{1}{12\pi}  
\int_0^{\infty} E_1 \, dE_1 \int_0^{\infty}  E_2 \, dE_2 
\int_0^{\pi}d\gamma\sin\gamma\cos^{2}\gamma\;
\varphi^{*}(E_1,\gamma)\varphi(E_2,\gamma)\;  \nonumber\\
& &  \left[  E^{2}_{1} + E^{2}_{2} + E_1 E_2 \right] 
J_0([E_1-E_2]\rho\,\sin\gamma)\; e^{i(E_2-E_1)t} \;,\cr \cr 
\langle {\hat T}^{03}(t,\rho)\rangle &=&
\frac{1}{12\pi}
\int_0^{\infty} E_1 \, dE_1 \int_0^{\infty}  E_2 \, dE_2 
\int_0^{\pi}d\gamma\sin\gamma\cos\gamma\;
\varphi^{*}(E_1,\gamma)\varphi(E_2,\gamma) \nonumber\\
& &  \left[  E^{2}_{1} + E^{2}_{2} + E_1 E_2 \right] 
\; J_0([E_1-E_2]\rho\,\sin\gamma)\; e^{i(E_2-E_1)t} \; .
\end{eqnarray}

Using rotational invariance around the $z$-axis we can express 
${\hat T}^{\alpha\beta}(x), {\hat T}^{3 \alpha}(x)$ and $ {\hat T}^{0
\alpha}(x),\alpha, \beta=1,2 $ as follows
\begin{eqnarray}
{\hat T}^{\alpha\beta}(x) &=& \delta^{\alpha\beta} \; a(t,\rho) +
e^{\alpha} \, e^{\beta}\; b(t,\rho) \cr \cr
 {\hat T}^{0\alpha}(x) &=& e^{\alpha} \,c(t,\rho) \quad , \quad 
 {\hat T}^{3 \alpha}(x)= e^{\alpha} \,d(t,\rho) \nonumber
\end{eqnarray}
where $ e^{\alpha} = (\cos\phi, \sin\phi)$. The coefficients $
a(t,\rho),  b(t,\rho), c(t,\rho)$ and $d(t,\rho)$ follow from the
calculation for $\rho = x$ (that is $\phi = 0$):
\begin{eqnarray}
a(t,\rho) &=& \langle {\hat T}^{22}(t,\rho=x)\rangle \quad , \quad 
  b(t,\rho) =  \langle {\hat T}^{11}(t,\rho=x)\rangle -  \langle {\hat
T}^{22}(t,\rho=x)\rangle \; ,\cr \cr
c(t,\rho) &=& \langle {\hat T}^{01}(t,\rho=x)\rangle \quad , \quad 
d(t,\rho)  = \langle {\hat T}^{13}(t,\rho=x)\rangle  \nonumber
\end{eqnarray}
We find from eqs.(\ref{paqcil}, \ref{tmassles2}) at $\rho = x$,
               
\begin{eqnarray}\label{cilsim2}
\langle {\hat T}^{11}(t,\rho=x)\rangle &=& \frac{1}{24\pi}
\int_0^{\infty} E_1 \, dE_1 \int_0^{\infty}  E_2 \,dE_2 
\int_0^{\pi}d\gamma\sin^{3}\gamma\;
\varphi^{*}(E_1,\gamma)\varphi(E_2,\gamma)\;  
e^{i(E_2-E_1)t} \nonumber\\
& &  \left[  E^{2}_{1} + E^{2}_{2} + E_1 E_2 \right]
\;\left[J_0([E_1-E_2]\rho\,\sin\gamma)-
J_2([E_1-E_2]\rho\,\sin\gamma)\right], \cr \cr
\langle {\hat T}^{22}(t,\rho=x)\rangle &=&
\frac{1}{12\pi}
\int_0^{\infty}  E_1 \,dE_1 \int_0^{\infty} E_2 \, dE_2 
\int_0^{\pi}d\gamma\sin^{2}\gamma\;
\varphi^{*}(E_1,\gamma)\varphi(E_2,\gamma) \nonumber\\
& &  \left[  E^{2}_{1} + E^{2}_{2} + E_1 E_2 \right]
\;J_1([E_1-E_2]\rho\,\sin\gamma)\;
\frac{e^{i(E_2-E_1)t}}{(E_1-E_2)\rho}\; , \cr \cr 
\langle {\hat T}^{01}(t,\rho=x)\rangle &=& \frac{i}{12\pi}
\int_0^{\infty}  E_1 \,dE_1 \int_0^{\infty} E_2 \,  dE_2 
\int_0^{\pi}d\gamma\sin^{2}\gamma\;
\varphi^{*}(E_1,\gamma)\varphi(E_2,\gamma) \nonumber\\
& &  \left[  E^{2}_{1} + E^{2}_{2} + E_1 E_2 \right]
\;J_1([E_1-E_2]\rho\,\sin\gamma)\; e^{i(E_2-E_1)t}\; , \cr \cr
 \langle {\hat T}^{13}(t,\rho=x)\rangle & = &
 \frac{i }{12\pi}  
\int_0^{\infty}    E_1 \,dE_1 \int_0^{\infty} E_2 \,  dE_2 
\int_0^{\pi}d\gamma\sin^{2}\gamma\cos\gamma\;
\varphi^{*}(E_1,\gamma)\varphi(E_2,\gamma)  \nonumber\\
& & \left[  E^{2}_{1} + E^{2}_{2} + E_1 E_2 \right]\;
J_1([E_1-E_2]\rho\,\sin\gamma)\; e^{i(E_2-E_1)t} \; ,
\end{eqnarray}
where $J_{n}(z)$ stand for Bessel functions of integer order. Cylindrical 
symmetry also results in
$$
\langle {\hat T}^{02}(t,\rho=x)\rangle = \langle {\hat
T}^{12}(t,\rho=x)\rangle =\langle {\hat T}^{23}(t,\rho=x)\rangle = 0.
$$

For a wave packet  symmetric  with respect to the $ x y $ plane,
$$
\varphi(E,\gamma) = \varphi(E,\pi - \gamma) \; ,
$$
and we find
$$
d(t,\rho)  = \langle {\hat T}^{13}(t,\rho=x)\rangle = 0 \; , \; 
 \langle {\hat T}^{03}(t,\rho)\rangle = 0 \; .
$$

Notice that 
$$
2 a(t,\rho) + b(t,\rho) +\langle {\hat T}^{33}(t,\rho)\rangle =
\langle {\hat T}^{00}(t,r)\rangle  
$$ 
due to the tracelessness of the energy-momentum tensor. 

In order to compute the asymptotic behaviour $\rho \to \infty$, 
we change in eq.(\ref{cilsim}-\ref{cilsim2}) the integration variables  
$$
E_2-E_1 = v \tau / \rho \quad , \quad E_2+E_1 = v\; .
$$
We find for the energy for a real $ \varphi(E, \gamma) $,
\begin{eqnarray}
\langle {\hat T}^{00}(t,\rho)\rangle &=& \frac{1}{64 \pi \; \rho}  \; 
\int_0^{\infty} v^5 \, dv \; \int_0^{\rho} d\tau \; \int_0^{\pi}
\sin\gamma \; d\gamma \; \varphi(\frac{v}{2}[1+\frac{\tau}{\rho}],\gamma )
\varphi(\frac{v}{2}[1-\frac{\tau}{\rho}],\gamma)\nonumber\\   
& & \cos(v\tau t/\rho ) \;
\left[1 - \frac23 {{\tau^2}\over {\rho^2}} 
- \frac13 {{\tau^4}\over {\rho^4}}\right]\; J_0(v\tau\sin\gamma)\; .
\nonumber
\end{eqnarray}
This representation is appropriate to compute the  limit
$\rho\rightarrow\infty$ with $t$ fixed. We find in such limit,
\begin{eqnarray}
\langle {\hat T}^{00}(t,\rho)\rangle & \buildrel{\rho\to \infty}\over=&
\displaystyle{ 
\frac{1}{2 \pi \; \rho}  \; \int_0^{\infty} E^4 dE \;  \int_{\sin\gamma >
t/\rho} \; d\gamma \; {{\varphi(E, \gamma)^2}\over{\sqrt{1 - \left({t
\over{\rho\sin\gamma}}\right)^2}}}} \; ,\nonumber
\end{eqnarray}
where we used the formula
$$
\int_0^{\infty} J_0(ax)\; \cos bx \; dx = {{\theta(a^2 - b^2)}\over
{\sqrt{a^2 - b^2}}} \; .
$$

We can further simplify $\langle {\hat T}^{00}(t,\rho)\rangle  $  for
$ \rho >> t $ yielding 
$$
\langle {\hat T}^{00}(t,\rho)\rangle  \buildrel{t\to \infty}\over=
\frac{1}{2 \pi \; \rho}  \; \int_0^{\infty} E^4 dE \;  \int_0^{\pi}
 \; d\gamma \; \varphi(E, \gamma)^2 + O({{t^2}\over {\rho^2}}) \; .
$$
That is, the energy decays as $ \rho^{-1} $ for large $  \rho $ and
fixed $ t $. 

We analogously compute the $\rho\rightarrow\infty, \; t  $ fixed
behaviour of the other components of $\langle {\hat
T}^{\mu\nu}(x)\rangle $  with the following results
\begin{eqnarray}
a(t,\rho) & \buildrel{\rho\to \infty}\over=&
\displaystyle{ 
\frac{1}{2 \pi \; \rho}  \; \int_0^{\infty} E^4 dE \;  \int_{\sin\gamma >
t/\rho} \; d\gamma \; \varphi(E, \gamma)^2\; \sin^2\gamma \;\sqrt{1 - \left({t
\over{\rho\sin\gamma}}\right)^2}} \; , \cr \cr
a(t,\rho) + b(t,\rho) & \buildrel{\rho\to \infty}\over=& 
\displaystyle{ 
\frac{t^2}{2 \pi \; \rho^3}  \; \int_0^{\infty} E^4 dE \;  \int_{\sin\gamma >
t/\rho} \; d\gamma \; {{\varphi(E, \gamma)^2}\over{\sqrt{1 - \left({t
\over{\rho\sin\gamma}}\right)^2}}}}  \cr \cr
 & \buildrel{\rho\to \infty}\over=& {{t^2}\over{ \rho^2}}\; \langle
{\hat T}^{00}(t,\rho)\rangle \; , \cr \cr
\langle {\hat T}^{33}(t,\rho)\rangle  &\buildrel{\rho\to \infty}\over=&
\displaystyle{ 
\frac{1}{2 \pi \; \rho}  \; \int_0^{\infty} E^4 dE \;  \int_{\sin\gamma >
t/\rho} \; d\gamma \cos^2\gamma\; {{\varphi(E, \gamma)^2}\over{\sqrt{1
- \left({t \over{\rho\sin\gamma}}\right)^2}}}} \; , \cr \cr
c(t,\rho) &\buildrel{\rho\to \infty}\over=&
\displaystyle{ 
\frac{t}{2 \pi \; \rho^2}  \; \int_0^{\infty} E^4 dE \;  \int_{\sin\gamma >
t/\rho} \; d\gamma \; {{\varphi(E, \gamma)^2}\over{\sqrt{1 - \left({t
\over{\rho\sin\gamma}}\right)^2}}}} \cr \cr
 &\buildrel{\rho\to \infty}\over=&
 {{t}\over{ \rho}}\; \langle {\hat T}^{00}(t,\rho)\rangle \; .\nonumber
\end{eqnarray}

\bigskip

The calculation of the $ t \to  \infty$ behaviour for  $\rho$ fixed can
be easily obtained from  eq.(\ref{cilsim}) undoing the integration
over $ \alpha $. That is,  
\begin{equation}\label{asiEt}
\langle {\hat T}^{00}(t,\rho)\rangle = { 1 \over {24 \pi^2}} \; 
\int_0^{\pi} d\gamma\sin\gamma\int_0^{2\pi}d\alpha 
\left[ f_3(\gamma, \alpha) f^{*}_1(\gamma, \alpha) +  f_1(\gamma,
\alpha) f^{*}_3(\gamma, \alpha) +  f_2(\gamma, \alpha) f^{*}_2(\gamma, \alpha) \right] \; ;
\end{equation}
where
\begin{equation}\label{fnc}
 f_n(\gamma, \alpha)\equiv \int_0^{\infty} E^n \; dE \; \varphi(E, \gamma)
\; e^{iE(\rho \sin\gamma\sin\alpha - t)} \; .
\end{equation}
We explicitly see here $ \langle {\hat T}^{00}(t,\rho)\rangle $ as a
superposition of  outgoing and ingoing  cylindrical waves.

The integral (\ref{fnc}) is dominated in the $ t \to \infty $ limit by
its lower bound. We find,
$$
 f_n(\gamma, \alpha) \buildrel{t\to \infty}\over= n! (-i)^{n+1} \;
\varphi(0, \gamma) \; t^{-n-1} \left[ 1 + O(t^{-1})\right] \; .
$$
Inserting this result in eq.(\ref{asiEt}) yields
$$
\langle {\hat T}^{00}(t,\rho)\rangle \buildrel{t\to \infty}\over=
- { 2 \over { 3 \pi \; t^6}}\; \int_0^{\pi}
d\gamma\; \sin\gamma\; \varphi(0, \gamma)^2 \left[ 1 + O(t^{-1})\right] \; .
$$

Using the same technique we find for the   $ t \to  \infty$ behaviour
for  $\rho$ fixed of the other $ \langle {\hat T}^{\mu \nu}(t,\rho)\rangle $
components,

\begin{eqnarray}
\langle {\hat T}^{33}(t,\rho)\rangle &\buildrel{t\to \infty}\over=&
- { 2 \over { 3 \pi \; t^6}}\; \int_0^{\pi}
d\gamma\; \sin\gamma\;\cos^2\gamma\; \varphi(0, \gamma)^2 \left[ 1 +
O(t^{-1})\right] \; , \cr \cr
a(t,\rho) &\buildrel{t\to \infty}\over=&
- { 1 \over { 3 \pi \; t^6}}\; \int_0^{\pi}
d\gamma\; \sin^3\gamma\; \varphi(0, \gamma)^2 \left[ 1 +
O(t^{-1})\right] \; , \cr \cr
b(t,\rho) &\buildrel{t\to \infty}\over=& 0 + O(t^{-7}) \; , \cr
c(t,\rho) &\buildrel{t\to \infty}\over=& 0 + O(t^{-7}) \; . \nonumber
\end{eqnarray}

As before, a typical form of the wave function $\varphi(E,\gamma)$
is  a gaussian Ansatz 
\begin{equation}\label{cilgau}
\varphi(E,\gamma) = \left(\frac{2c}{\pi}\right)^{3/2} 
\; e^{- E^2 (a^2 \sin^2\gamma + b^2 \cos^2\gamma)}
\end{equation}
where $ a^2 $ and $ b^2 $ are constants.

{F}rom the expressions above we can see that 
for large $t$ and $\rho$ fixed the energy density decays 
rapidly as $O(1/t^6)$ whilst the energy flux vanishes.\ 
For $\rho$ large and fixed $t$ the 
situation is similar to the spherical one, the energy density  decays
as $1/\rho$ and the energy flux decays as $1/\rho^2$.

If we go now to regions where both $\rho$ and $t$ are 
small, we find after a careful examination of
$\langle{\hat T}^{\mu\nu}\rangle$ that  {\bf all} of its components are
finite constants in such limit.\ 

\section{Final remarks.}

An important point we need to stress, 
is the fact that, in contrast to what happens in the
classical theory where there is a logarithmic divergence as we approach
the core of the string \cite{edal}, in the quantum theory we can
obtain a metric  where, if any
divergences are present these are not related at all to the position of
the string \cite{alfredo}.\ We showed, that the localised behaviour of the 
string given
by $\delta(x-X(\sigma,\tau))$ dissappears when we take into account the
quantum nature of the strings.\ 
 In the {\em classical theory of radiating strings},
divergences coincide with the source of the gravitational field. However,
as we have seen, quantum mechanically this is not the case. 
The source position is smeared by the quantum fluctuations. The probability 
amplitude for the centre of mass string position
is given by the Fourier transform of $\varphi ({\vec p})$ and is a
smooth function peaked at the origin.

We show that  {\bf all}  components of $\langle{\hat
T}^{\mu\nu}\rangle$ are finite everywhere  both in the
 cylindrically and in the  spherically symmetric cases. 
[This finiteness was already observed in ref.\cite{tmunu} for
shock-wave spacetimes].  The present results support the belief that
string theory is a finite theory.

\section*{Acknowledgements}
E.J.C. and A.V. would like to thanks D. Bailin for his  useful comments.
\pagebreak

\end{document}